\documentclass[showpacs,twocolumn,preprintnumbers,showkeys,superscriptaddress,amsmath,amssymb,nofootinbib]{revtex4-1}

\usepackage{lipsum}

\usepackage{bbold}
\usepackage{color}
\usepackage{latexsym}
\usepackage{amsmath}
\usepackage{amssymb}
\usepackage[utf8]{inputenc}
\usepackage{amsfonts}
\usepackage{bm}
\usepackage{bbold}

\usepackage{eufrak}
\usepackage{euscript}
\usepackage{graphics}
\usepackage{graphicx}

\newcommand{\be}{\begin{equation}}
\newcommand{\ee}{\end{equation}}
\newcommand{\ba}{\begin{eqnarray}}
\newcommand{\ea}{\end{eqnarray}}

\begin{document}

%\linenumbers
%
%\begin{center}

\title{\Large Non-local material vacuum and Cherenkov radiation
in non-linear massive $3$D-Electrodynamics}

\author{Patricio Gaete} \email{patricio.gaete@usm.cl}
\affiliation{Departamento de F\'{i}sica and Centro Cient\'{i}fico-Tecnol\'ogico de Valpara\'{i}so-CCTVal,
Universidad T\'{e}cnica Federico Santa Mar\'{i}a, Valpara\'{i}so, Chile}

\author{J. A. Helay\"el-Neto}\email{helayel@cbpf.br}
\affiliation{Centro Brasileiro de Pesquisas F\'isicas, Rua Dr. Xavier Sigaud
150, Urca, Rio de Janeiro, Brasil, CEP 22290-180}

\date{\today}

\begin{abstract}
We examine the effects of electromagnetic field non-linearities in $3$ space-time dimensions. We focus on how these non-linearities influence permittivity and susceptibility. This, in turn, leads to changes in the refractive index through the use of the dispersion relation in the context of massless and massive non-linear electrodynamics. We also verify that, by inspecting the model addressed in the frequency/wave vector space, we identify the characteristics of a non-local material in the behavior of the vacuum, which exhibits a spatially-dispersive profile. Furthermore, it is important to highlight that the cause of this phenomenon is the de Broglie-Proca mass term.
We subsequently investigate the electromagnetic radiation emitted by a moving charged particle interacting with a medium for massless and massive non-linear electrodynamics.
Our findings indicate that the radiation is driven by the medium through which the particle travels, similar to what is observed in the Cherenkov effect.
\end{abstract}

\maketitle

\pagestyle{myheadings}
\markright{Non-local material vacuum and Cherenkov radiation
in non-linear massive $3$D-Electrodynamics}

\section{Introduction}

The study of gauge theories in (2+1)-dimensions and their physical implications, such as massive gauge fields and fractional statistics, has been extensively explored in the literature \cite{Deser,Dunne,Khare,Banerjee}. These theories are particularly intriguing due to their peculiar dynamical structure and relevance to condensed matter systems. An example that illustrates the above properties is the three-dimensional Chern-Simons gauge theory, as it allows the implementation of the charge-flux composite model of anyons \cite{Wilczek1,Wilczek2}. The motivation for researching three-dimensional theories stems from the fact that three-dimensional Yang-Mills theories are super-renormalizable \cite{Deser}, and the presence of mass in the gauge fields does not violate gauge symmetry. Additionally, topologically massive Yang-Mills theories exhibit ultraviolet finiteness \cite{Cima,Barnich}. As mentioned, three-dimensional gauge theories are instrumental in investigating low-dimensional condensed matter systems. By using effective gauge theories, they efficiently describe collective excitations, such as spin or pairing fluctuations. Furthermore, these theories are relevant to high-$T_{C}$ superconductivity, where planarity is often a reliable approximation 
\cite{Khveshchenko}. It is also important to note that three-dimensional Yang-Mills theories facilitate accurate comparisons between continuum and lattice calculations 
\cite{Feuchter}.

On the other hand, vacuum polarization in quantum electrodynamics (QED) is a well-known phenomenon resulting from the polarization of virtual electron-positron pairs. This effect leads to non-linear interactions between electromagnetic fields and has been a fascinating research topic since it was first discovered \cite{Euler, Adler, Costantini, Biswas, Tommasini, Ferrando}. A key issue in this context is the scattering of photons by photons, which has significant implications, including vacuum birefringence and vacuum dichroism. This topic has been widely explored from various perspectives. Meanwhile, despite considerable advancements in research, experimental confirmation of this prediction is still elusive 
\cite{Bamber, Burke, Pike}. Interestingly, it should be recalled here that the ATLAS and CMS collaborations at the Large Hadron Collider (LHC) recently reported the emission of high-energy gamma-gamma pairs resulting from virtual gamma-gamma scattering in ultraperipheral Pb-Pb collisions \cite{Atlas, Cms}. While these findings are remarkable, it is important to note, as mentioned in \cite{Robertson2}, that no changes have been observed in the optical properties of the vacuum. Moreover, with the advent of laser facilities, there have been exciting proposals to investigate quantum vacuum nonlinearities \cite{Battesti, Ataman}. One appealing experiment is the DeLLight project \cite{Robertson2}, which exploits the alteration in the refractive index stemming from nonlinear electrodynamics. 

With these ideas in mind, in previous works \cite{Rad1, Rad2, Rad3}, we have examined the radiation phenomenon associated with various models of $(3+1)$-dimensional nonlinear electrodynamics. Specifically, our investigation has focused on the impact of the new medium on radiation production. To do this, we calculated the radiated energy by employing the conventional approach of determining the Poynting vector. Our findings show the crucial role vacuum electromagnetic nonlinearities play in triggering the radiated power in $(3+1)$-dimensions.

At this point, it is essential to highlight a noteworthy study on Cherenkov radiation in $(2+1)$ dimensions, as discussed in \cite{Gerlach}. In fact, \cite{Gerlach} presents the first observation of Cherenkov surface waves, where free electrons emit radiation within a two-dimensional framework.  

Inspired by the preceding considerations, the purpose of this discussion is to further elaborate on the physical impact of the medium on the generation of electromagnetic radiation, specifically in the context of massless and massive non-linear electrodynamics in $(2+1)$-dimensions. For this purpose, as in the case of $(3+1)$-dimensions, we will calculate the radiated energy by following the conventional approach of determining the Poynting vector. As will be shown, our analysis reveals the crucial role that vacuum electromagnetic non-linearities play in triggering this radiated energy. At this point, we should mention the emission of electromagnetic radiation by moving charges in a Maxwell-Chern-Simons vacuum, which displays the properties of Cherenkov radiation 
\cite{Lehnert}.

Finally, an interesting aspect worth highlighting is that, by virtue of a non-vanishing de Broglie-Proca photon mass, when the model is examined in the frequency/wave vector space, we make use of the special mapping between the curl and the divergence differential operators, specific to two-dimensional Euclidean space, to demonstrate the possibility of defining an electric permittivity tensor that depends both on the frequency and the wave vector. This phenomenon is known as temporal and spatial dispersions, respectively \cite{Pekkar}. This then means that the vacuum of the three-dimensional de Broglie-Proca massive electrodynamics behaves as a non-local material. According to the papers of Refs. \cite{Mikki1,Mikki2}, non-local materials may be suitably engineered to achieve unique electromagnetic performance, enabling them to function as planar non-local metamaterials with potential applications in nano-electromagnetic technology.

Our work is organized according to the following outline: In Sec. 2, we focus on how these non-linearities influence permittivity and susceptibility, leading to changes in the refractive index through the dispersion relation. In Sec. 3, we consider the calculation of electromagnetic radiation. Finally, some concluding remarks are presented in Sec. 4. 
 
In our conventions, the signature of the metric is ($+1,-1,-1$). 

\section{General aspects}

We will start by briefly introducing an extension of our earlier formalism \cite{Rad1,Rad2,Rad3} to include non-linear electrodynamics (NLED) in $(2+1)$-dimensions. In pursuit of this objective, we will introduce the most general Lagrangian density ${\cal L}$. This Lagrangian density is a function of the Lorentz-invariant and gauge-invariant bilinear, ${\cal F}$,
\begin{equation}
{\cal F} =  - \frac{1}{4}F_{\mu \nu }^2 = \frac{1}{2}({{\bf E}^2} - {B^2}), \label{Fieldeq05}
\end{equation}
with ${F^{\mu \nu }} = {\partial ^\mu }{A^\nu } - {\partial ^\nu }{A^\mu }$. Additionally, it is important to recall that in $(2+1)$-dimensions, the magnetic field is a scalar field.

Next, as was explained in \cite{Rad1,Rad2,Rad3}, we express the potential $A^{\mu}$ as the sum of $a^{\mu}$ and $A_{B}^{\mu}$, where $a^{\mu}$ represents the photon field and $A_{B}^{\mu}$ represents the background field. This decomposition allows us to represent the tensor $F^{\mu\nu}$ as ${F^{\mu \nu }} = {f^{\mu \nu }} + F_B^{\mu \nu}$, where ${f^{\mu \nu }}$ is the electromagnetic field strength tensor of the propagating excitation and $F_B^{\mu \nu}$ corresponds to the electromagnetic field strength tensor of the background fields. We further note that, in our development, these background fields do not depend on the space-time coordinates. With this information at hand, we now proceed to expand the Lagrangian density ${\cal L}({\cal F})$ around the background fields while keeping terms up to the second order in the propagating field, namely,
\begin{eqnarray}
{{\cal L}^{\left( 2 \right)}} &=&  - \frac{1}{4}{C_1}{f^{\mu \nu }}{f_{\mu \nu }} + \frac{1}{8}k_B^{\mu \nu \kappa \lambda }{f_{\mu \nu }}{f_{\kappa \lambda }} \nonumber\\
&+& \frac{1}{2}\,m\,{\varepsilon ^{\mu \nu \kappa }}{a_\mu }{\partial _\nu }{a_\kappa } 
+ \frac{1}{2}{M^2}{A_\mu }{A^\mu }, \label{Fieldeq10}
\end{eqnarray}
in the above, we have defined the background tensor as 
\begin{equation}
k_B^{\mu \nu \kappa \lambda } = {D_1}F_B^{\mu \nu }F_B^{\kappa \lambda }. \label{Fieldeq15}
\end{equation}
Our general Lagrangian density incorporates a topological mass by including the Chern-Simons term and a mass parameter denoted by $M$ introduced via a Proca term.

Additionally, we emphasize that the coefficients $C_{1}$ and $D_{1}$ should be evaluated at the background fields ${\bf E}$ and $B$:
\begin{eqnarray}
{C_1} = {\left. {\frac{{\partial {\cal L}}}{{\partial {\cal F}}}} \right|_{{\bf E},B}}\, , \,{D_1} = {\left. {\frac{{{\partial ^2}{\cal L}}}{{\partial {{\cal F}^2}}}} \right|_{{\bf E},B}}. \label{Fieldeq20}
\end{eqnarray}

From the Lagrangian density (\ref{Fieldeq10}), we can derive the corresponding field equations  
\begin{equation}
{C_1}{\partial _\mu }{f^{\mu \nu }} - \frac{1}{2}k_B^{\mu \nu \kappa \lambda }{\partial _\mu }{f_{\kappa \lambda }} + m{\tilde f^\nu } + {M^2}{a^\nu } = 0, \label{Fieldeq25a}
\end{equation}
together with the subsidiary condition
\begin{equation}
{\partial ^\nu }{a_\nu } = 0, \label{Fieldeq25b}
\end{equation}
where ${\tilde f^\mu } = \frac{1}{2}{\varepsilon ^{\mu \nu \kappa }}{f_{\nu \kappa }}$. Note that ${\tilde f^0} = {\tilde f_0} =  - b$, ${\tilde f^i} =  - {\tilde f_i} =  - {\varepsilon _{ij}}{e_j}$, ${f_{0i}} = {e_i}$ and ${f_{ij}} =  - {\varepsilon _{ij}}b$.
It is also important to observe that, for a purely magnetic background, we have $k_B^{ijkl} = {D_1}F_B^{ij}F_B^{kl} = {D_1}{\varepsilon _{ij}}{\varepsilon _{kl}}{B^2}$. Accordingly, the field equations can be brought to the form 
\begin{subequations}
\begin{eqnarray}
{C_1}\,\nabla  \cdot {\bf e} - m \,b + M^{2} \varphi = 0, \label{Fieldeq30} \\
\left( {{C_1} - {D_1}{B^2}} \right)\tilde \nabla b - m\,\tilde {\bf e} + {M^2}\,{\bf a} = {C_1}\frac{{\partial \,{\bf e}}}{{\partial t}}, \label{Fieldeq35} \\
\nabla  \times {\bf e} =  - \frac{\partial \,b}{{\partial t}}. \label{Fieldeq40}
\end{eqnarray}
\end{subequations}
In the above equations, we have defined $\varphi=a_{0}$, ${\tilde e_i} = \varepsilon _{ij}^{LC}{e_j}$ and 
${\tilde \nabla _i} = \varepsilon _{ij}^{LC}{\partial _j}$. In passing, we note that the superscript LC on 
$\varepsilon _{ij}^{LC}$  is to emphasize that this is the Levi-Civita tensor.

It is also important to observe that, to ascertain the permittivity and permeability of the newly defined vacuum electromagnetic, we will decompose the electromagnetic fields ${\bf e}$ and $b$ into plane wave components; that is,
\begin{subequations}
\begin{eqnarray}
\varphi  = {\varphi _0}\,{e^{i\left( {{\bf k} \cdot {\bf x} - \omega t} \right)}}, \,\,\,{\bf a} = {{\bf a}_0}\,{e^{i\left( {{\bf k} \cdot {\bf x} - \omega t} \right)}},  \label{Fieldeq45} \\
{\bf e} = {{\bf e}_0}\,{e^{i\left( {{\bf k} \cdot {\bf x} - \omega t} \right)}}, \,\,\,b = {b_0}\,{e^{i\left( {{\bf k} \cdot {\bf x} - \omega t} \right)}}.  \label{Fieldeq50}
\end{eqnarray}
\end{subequations}
From equation (\ref{Fieldeq30}) we readily deduce that
\begin{equation}
{\varphi _0} = \frac{{ - i{C_1}{\bf k} \cdot {{\bf e}_0} + m\,{b_0}}}{{{M^2}}}, \label{Fieldeq55}
\end{equation}
with ${M^2} \ne 0$.
Utilizing the definition of electric field, expressed as  ${{\bf e}_0} =  - i\,{\bf k}\,\varphi  + i\,\omega \,{{\bf a}_0}$, we thus find    
\begin{equation}
{{\bf a}_0} =  - \frac{i}{\omega }{{\bf e}_0} - i\frac{{{C_1}}}{{\omega {M^2}}}{\bf k}\left( {{\bf k} \cdot {{\bf e}_0}} \right) + \frac{m}{{\omega {M^2}}}{\bf k}\,{b_0}. \label{Fieldeq60}
\end{equation}

Making use of the foregoing results, we finally obtain from equation  (\ref{Fieldeq35}) the following field equation
\begin{equation}
\tilde \nabla h = \frac{\partial }{{\partial t}}\,{\bf d}. \label{Fieldeq65}
\end{equation}
To arrive at the last equation, we employed ${d_i} = {\varepsilon _{ij}}{e_j}$ and 
$h = \frac{b}{\mu }$. Consequently, this implies that the permittivity, ${\varepsilon _{ij}}$, and the permeability, $\mu$, are defined as follows:
\begin{equation}
{\varepsilon _{ij}} \equiv \left( {{C_1} - \frac{{{M^2}}}{{{\omega ^2}}}} \right){\delta _{ij}} - \frac{{{C_1}}}{{{\omega ^2}}}{k_i}{k_j} + i\frac{m}{{{\omega ^3}}}{k_i}{\tilde k_j} + i\frac{m}{\omega }\varepsilon _{ij}^{LC}, \label{Fieldeq70}
\end{equation}
and
\begin{equation}
\mu  = \frac{1}{{\left( {{C_1} - {D_1}{B^2}} \right)}}. \label{Fieldeq75}
\end{equation}

 The electric permittivity tensor presented in Eq. (\ref{Fieldeq70}) exhibits two contributions that depend on the components of the wave vector, thereby demonstrating a spatially-dispersive behavior. Our derivation shows that the de Broglie-Proca mass term, denoted by $M^2$, is the source of this phenomenon. Eq. (\ref{Fieldeq55}) serves as the starting point for the derivation that leads to Eq. (\ref{Fieldeq70}). It is important to notice that our approach to this equation is only valid when the parameter $M^2$ is non-vanishing. Though this parameter does not appear explicitly in both terms with {\bf k}-dependence, it is crucial that $M^2$ be non-zero in the intermediate steps. We have verified that in the case the mass is purely topological (Chern-Simons mass term), the wave-vector dependence of the dielectric tensor does not arise. This characterizes the vacuum of the full-fledged model we are investigating as a non-local material, which, according to the works of Ref. \cite{Mikki3}, may be specially engineered to behave as a nonlocal metamaterial. This may be an interesting aspect to exploit in a future project; however, as far as this contribution is concerned, this is out of our scope.

Next, the field equations provide the necessary information to explore another aspect of vacuum electromagnetic properties. We intend to understand how the parameters and external fields from the nonlinear (NL) sector are connected to the $m$ and $M$ parameters. In the following subsection, we will thoroughly examine this connection.\\

\subsection{Dispersion relations}

To address this issue, exploring the dispersion relations (DRs) associated with an electromagnetic wave propagating in an external electromagnetic background is essential. To achieve this, we will again decompose the electromagnetic fields ${\bf e}$ and $b$ into plane waves. Consequently, by starting from equation (\ref{Fieldeq35}), it follows that 
\begin{widetext}
\begin{eqnarray}
\left( {{C_1} - {D_1}{B^2}} \right)i{{\tilde k}_i}{b_0} = \left( { - i\omega } \right)\left[ {\left( {{C_1} - \frac{{{M^2}}}{{{\omega ^2}}}} \right){\delta _{ij}} - \frac{{{C_1}}}{{{\omega ^2}}}{k_i}{k_j} + i\frac{m}{{{\omega ^3}}}{k_i}{{\tilde k}_j} + i\frac{m}{\omega }\varepsilon _{ij}^{LC}} \right]{e_{oj}}, \label{Fieldeq80}
\end{eqnarray}
\end{widetext}
and making use of equation (\ref{Fieldeq40}), we thus have
\begin{equation}
M_{ij}e_j=0,  \label{Fieldeq80}
\end{equation}
where the matrix $M_{ij}$ is given by
\begin{eqnarray}
{M_{ij}} &=& \left[ {{C_1}\left( {{\omega ^2} - {{\bf k}^2}} \right) + {D_1}{B^2}{{\bf k}^2} - {M^2}} \right]{\delta _{ij}} \nonumber\\
 &-& \left( {{C_1} + {D_1}{B^2}} \right){k_i}{k_j} + i\frac{m}{\omega }{k_i}{{\tilde k}_j} + im\omega \varepsilon _{ij}^{LC}. \label{Fieldeq85}
\end{eqnarray}
with ${\tilde k_j} =  - {\varepsilon _{mj}}{k_m}$. The previous matrix is of the type 
${M_{ij}} = a{\delta _{ij}} + b{k_i}{k_j} + c{k_i}{\tilde k_j} + d\varepsilon _{ij}^{LC}$,
whose determinant is given by $\det M = {a^2} + {d^2} + \left( {ab - cd} \right){{\bf k}^2}$.
From the condition  $\det M=0$, we end up with the following dispersion relation
\begin{eqnarray}
{\left[ {{C_1}\left( {{\omega ^2} - {{\bf k}^2}} \right) + {D_1}{B^2}{{\bf k}^2} - {M^2}} \right]^2}  \nonumber\\
 - \left[ {{C_1}\left( {{\omega ^2} - {{\bf k}^2}} \right) + {D_1}{B^2}{{\bf k}^2} - {M^2}} \right]\left( {{C_1} + {D_1}{B^2}} \right){{\bf k}^2} \nonumber\\
 - {m^2}\left( {{\omega ^2} - {{\bf k}^2}} \right) = 0. \nonumber\\
\label{Fieldeq90}
\end{eqnarray}

It is of interest also to notice that by defining $\Omega  \equiv {\omega ^2} - {{\bf k}^2}$, $\Lambda  \equiv {D_1}{B^2}{{\bf k}^2} - {M^2}     $, $\Gamma  \equiv \left( {{C_1} + {D_1}{B^2}} \right){{\bf k}^2}$, equation (\ref{Fieldeq90}) can be brought to the form
\begin{equation}
C_1^2{\Omega ^2} + \left( {2{C_1}\Lambda  - {C_1}\Gamma  - {m^2}} \right)\Omega  + \Lambda \left( {\Lambda  - \Gamma } \right) = 0. \label{Fieldeq95}
\end{equation}

From this last expression, we can proceed to obtain some special cases:
\begin{itemize}
\item Maxwell-Chern-Simons.
In this case $C_{1}=1$, $\Lambda=0$, $\Gamma={\bf k}^{2}$ and $M^{2}=0$.
Thus we find
\begin{equation}
\Omega \left( {\Omega  - {m^2}} \right) = 0. \label{Fieldeq100}
\end{equation}
Accordingly, we have ${\omega ^2} = {{\bf k}^2}$ and ${\omega ^2} = {{\bf k}^2} + {m^2}$.
\item Maxwell with a Proca term.
Here, $C_{1}=1$, $\Lambda=-M^{2}$, $\Gamma={\bf k}^{2}$ and $m=0$.
Consequently,
\begin{equation}
{\Omega ^2} - \left( {{{\bf k}^2} + 2{M^2}} \right)\Omega  + {M^2}\left( {{{\bf k}^2} + {M^2}} \right) = 0. \label{Fieldeq105}
\end{equation}
In a reference system at rest, ${\bf k}={\bf 0}$, one encounters the Proca photon mass $\omega^{2}=M^{2}$.
\item Non-linear Maxwell-Chern-Simons.
In this case, $\Lambda  = {D_1}{B^2}{{\bf k}^2}$, $\Lambda  - \Gamma  =  - {C_1}{{\bf k}^2}$.
With such a choice, the dispersion relation is given by 
\begin{equation}
C_1^2{\Omega ^2} + \left( {{C_1}{D_1}{{\bf k}^2} - C_1^2{{\bf k}^2} - {m^2}} \right)\Omega  - {C_1}{D_1}{B^2}{{\bf k}^4} = 0. \label{Fieldeq110}
\end{equation}
Again, in a reference system at rest, we obtain ${\omega ^2} = \frac{{{m^2}}}{{C_1^2}}$. This means that the non-linearity modules the Chern-Simons photon mass.
\end{itemize}

Finally, making use of equation (\ref{Fieldeq95}), we can further obtain the refractive index for non-linear electrodynamics with a Proca term, namely, 
\begin{equation}
{n^2} = \frac{{{C_1} - \frac{{{M^2}}}{{{\omega ^2}}}}}{{\left( {{C_1} - {D_1}{B^2}} \right)}}. \label{Fieldeq115}
\end{equation}

It is to be specially noted that, in the case of pure non-linear Logarithmic electrodynamics, 
\begin{equation}
{\cal L} =  - {\beta ^2}\ln \left[ {1 - \frac{{\cal F}}{{{\beta ^2}}}} \right], \label{Fieldeq120}
\end{equation}
the refractive index becomes
\begin{equation}
{n^2} = \frac{{\left( {1 + \frac{{{B^2}}}{{2{\beta ^2}}}} \right)}}{{\left( {1 - \frac{{{B^2}}}{{2{\beta ^2}}}} \right)}}. \label{Fieldeq125}
\end{equation}

With the newly defined electromagnetic vacuum characterized by the refractive index (\ref{Fieldeq115}), we can now explore aspects of electromagnetic radiation in the following section.

\section{Electromagnetic radiation}

As already mentioned, our immediate objective is to calculate the electromagnetic radiation produced by a moving charged particle interacting with a medium characterized by an electromagnetic vacuum resulting from massless and massive nonlinear electrodynamics. We will examine these calculations separately. We draw attention to the fact that proceeding in this manner is motivated by the observation that when we take the mass limit approaching zero, one does not recover the massless case.

\subsection{Massless nonlinear electrodynamics}

To put into context our discussion, let us begin by examining the Maxwell equations for a moving charged particle in a medium characterized by massless, nonlinear electrodynamics:
\begin{eqnarray}
\nabla  \cdot {\bf d} = 2\pi {\rho _{ext}}, \nonumber\\
 {\tilde \nabla}  \cdot {\bf e} = \frac{1}{c}\frac{{\partial b}}{{\partial t}}, \nonumber\\
{\tilde \nabla}b = \frac{{\mathord{\buildrel{\lower3pt\hbox{$\scriptscriptstyle\leftrightarrow$}} 
\over \varepsilon } \mu }}{c}\frac{{\partial \,{\bf e}}}{{\partial t}} + \frac{{2\pi \mu }}{c}{{\bf J}_{ext}}. \label{RadLog05}
\end{eqnarray}
 In the above, $\rho_{ext}$ represents the external charge density, and ${\bf J}_{ext}$ represents the external current density. Also, ${\bf d} = \mathord{\buildrel{\lower3pt\hbox{$\scriptscriptstyle\leftrightarrow$}} 
\over \varepsilon } {\bf e}$, $h = {\mu ^{ - 1}}b$, ${\varepsilon _{ij}} = i{C_1}{\delta _{ij}}$ and $\mu  =  - \frac{i}{{{C_1} - {D_1}{B^2}}}$.

We further observe that the equation for the electric field can also be expressed in the following form:
\begin{eqnarray}
\left( {{\nabla ^2} - \frac{{\mathord{\buildrel{\lower3pt\hbox{$\scriptscriptstyle\leftrightarrow$}} 
\over \varepsilon } \mu }}{{{c^2}}}\frac{{{\partial ^2}}}{{\partial {t^2}}}} \right){\bf e} = 2\pi {\mathord{\buildrel{\lower3pt\hbox{$\scriptscriptstyle\leftrightarrow$}} 
\over \varepsilon } ^{ - 1}}\nabla {\rho _{ext}} + \frac{{2\pi }}{{{c^2}}}\mu \frac{{\partial {{\bf J}_{ext}}}}{{\partial t}}, \label{RadLog10}
\end{eqnarray}
where the given external charge and current densities are: ${\rho _{ext}} = 2\pi Q\delta \left( {\omega  - {k_y}v} \right)$ and ${{\bf J}_{ext}} = 2\pi Qv\delta \left( {\omega  - {k_y}v} \right)\pmb {\hat e_y}$. For simplicity, we assume that the $y$-axis is the direction of the moving charged particle.

Next, as we mentioned in \cite{Rad1,Rad2,Rad3}, we perform a Fourier transform to momentum space via
\begin{equation}
F(t,{\bf x}) = \int {\frac{{d \omega {d^2}{\bf k}}}{{{{\left( {2\pi } \right)}^3}}}} {e^{ - i \omega t + i{\bf k} \cdot {\bf x}}}F\left( {\omega,{\bf k}} \right), \label{RadLog15}
\end{equation}
where $F$ stands for the electric field. 
Once this is completed, we obtain the following expression for the electric field: 
\begin{equation}
\left( {{k^2} - \frac{{\mathord{\buildrel{\lower3pt\hbox{$\scriptscriptstyle\leftrightarrow$}} 
\over \varepsilon } \mu }}{{{c^2}}}{\omega ^2}} \right){\bf e}\left( {\omega ,k} \right) =  - i2\pi\, \mathord{\buildrel{\lower3pt\hbox{$\scriptscriptstyle\leftrightarrow$}} 
\over \varepsilon }^{-1} \,{\bf k}\,{\rho _{ext}} + i\frac{{2\pi }}{{{c^2}}}\mu \,\omega \,{{\bf J}_{ext}}. \label{RadLog20}
\end{equation}
Similarly, in Fourier space, the external charge and current densities take the form: ${\rho _{ext}} = 2\pi Q\delta \left( {\omega  - {k_y}v} \right)$ and ${{\bf J}_{ext}} = 2\pi Qv\delta \left( {\omega  - {k_y}v} \right) \pmb {\hat e_y}$.

By following the same procedure as in \cite{Rad1,Rad2,Rad3}, we will now calculate ${\bf e}\left( {w,{\bf x}} \right)$. In this case, ${\bf e}\left( {w,{\bf x}} \right)$ is expressed as
\begin{equation}
{\bf e}\left( {\omega ,{\bf x}} \right) = \frac{1}{{4{\pi ^2}}}\int_{ - \infty }^\infty  {d{k_x}d{k_y}} {e^{i{\bf k} \cdot {\bf x}}}{\bf e}\left( {\omega ,{\bf k}} \right). \label{RadLog25}
\end{equation}
Next, we note that the electric field can be expressed as
\begin{equation}
{e_k} =  - i2\pi M_{ki}^{ - 1}\,\mathord{\buildrel{\lower3pt\hbox{$\scriptscriptstyle\leftrightarrow$}} 
\over \varepsilon } _{ij}^{ - 1}\,{k_j}\,{\rho _{ext}} + i\frac{{2\pi }}{{{c^2}}}\mu\, \omega\, M_{ki}^{ - 1}\,J_i^{ext}, \label{RadLog30}
\end{equation}
where $M_{ki}^{ - 1} = {\left( {{{\bf k}^2} - {n^2}\frac{{{\omega ^2}}}{{{c^2}}}} \right)^{ - 1}}{\delta _{ki}}$ and $\mathord{\buildrel{\lower3pt\hbox{$\scriptscriptstyle\leftrightarrow$}} 
\over \varepsilon } _{ij}^{ - 1} =  - iC_1^{ - 1}{\delta _{ij}}$.
As a consequence, the electric field components take the following forms:
\begin{eqnarray}
{e_1}\left( {\omega ,x} \right) &=&  - \frac{Q}{{v{C_1}}}\int_{-\infty}^{\infty}  {d{k_x}d{k_y}} \left( {{k_x} + {k_y}} \right)\frac{1}{{\left( {{{\bf k}^2} - {n^2}\frac{{{\omega ^2}}}{{{c^2}}}} \right)}} \nonumber\\
&\times&\delta \left( {{k_y} - {\omega  \mathord{\left/
 {\vphantom {\omega  v}} \right.
 \kern-\nulldelimiterspace} v}} \right){e^{i\left( {{k_x}x + {k_y}y} \right)}}, \label{RadLog35a}
\end{eqnarray}
and
\begin{eqnarray}
{e_2}\left( {\omega ,x} \right) &=&  - Q\int_{-\infty}^{\infty}  {d{k_x}d{k_y}} \left( {\frac{{{k_y}}}{{v{C_1}}} - \frac{{i\mu \omega }}{{{c^2}}}} \right)\frac{1}{{\left( {{{\bf k}^2} - {n^2}\frac{{{\omega ^2}}}{{{c^2}}}} \right)}} \nonumber\\
&\times&\delta \left( {{k_y} - {\omega  \mathord{\left/
 {\vphantom {\omega  v}} \right.
 \kern-\nulldelimiterspace} v}} \right){e^{i\left( {{k_x}x + {k_y}y} \right)}}. \label{RadLog35b}
\end{eqnarray}

When we integrate over $k_{y}$, the electric field components become:
\begin{eqnarray}
{e_1}\left( {\omega ,x} \right) &=&  - \frac{Q}{{v{C_1}}}{e^{iy{\omega  \mathord{\left/
 {\vphantom {\omega  v}} \right.
 \kern-\nulldelimiterspace} v}}}\int_{-\infty}^{\infty}  {d{k_x}} {e^{i{k_x}x}}  \nonumber\\
&\times&\left[ {\frac{{{k_x}}}{{\left( {k_x^2 - {\sigma ^2}} \right)}} + \frac{\omega }{v}\frac{1}{{\left( {k_x^2 - {\sigma ^2}} \right)}}} \right], \label{RadLog40a}
\end{eqnarray}
and
\begin{eqnarray}
{e_2}\left( {\omega ,x} \right) &=& \frac{Q}{{{C_1}}}{e^{iy{\omega  \mathord{\left/
 {\vphantom {\omega  v}} \right.
 \kern-\nulldelimiterspace} v}}}\frac{{\omega {n^2}}}{{{c^2}}}\left( {1 - \frac{{{c^2}}}{{{n^2}{v^2}}}} \right)\int_{-\infty}^{\infty}  {d{k_x}} {e^{i{k_x}x}} \nonumber\\
&\times&\frac{1}{{\left( {k_x^2 - {\sigma ^2}} \right)}}, \label{RadLog40b}
\end{eqnarray}
where ${\sigma ^2} = \frac{{{n^2}{\omega ^2}}}{{{c^2}}}\left( {1 - \frac{{{c^2}}}{{{n^2}{v^2}}}} \right)$.

Then, by integrating over $k_{x}$ and making further manipulations, we find that the electric field can be brought to the form 
\begin{eqnarray}
{\bf e}\left( {\omega ,{\bf x}} \right) &=& \frac{{{\pi ^{{3 \mathord{\left/
 {\vphantom {3 2}} \right.
 \kern-\nulldelimiterspace} 2}}}}}{{\sqrt 2 }}\frac{Q}{{{C_1}}}{e^{iy{\omega  \mathord{\left/
 {\vphantom {\omega  v}} \right.
 \kern-\nulldelimiterspace} v}}}\frac{{\sqrt x }}{{\sqrt \sigma  }} \nonumber\\
&\times&\!\!
\left[ {\frac{1}{v}\left( {\sigma  + \frac{\omega }{v}} \right) \pmb {{\hat e}_x} - \frac{{\omega {n^2}}}{{{c^2}}}\left( {1 - \frac{{{c^2}}}{{{n^2}{v^2}}}} \right) \pmb {{\hat e}_y}} \right]\nonumber\\
&\times&H_{{1 \mathord{\left/
 {\vphantom {1 2}} \right.
 \kern-\nulldelimiterspace} 2}}^{\left( 1 \right)}\left( {\sigma x} \right). \label{RadLog45}
 \end{eqnarray}
Here, $H_{{1 \mathord{\left/{\vphantom {1 2}} \right.
 \kern-\nulldelimiterspace} 2}}^{(1)}\left( {\sigma x} \right)$ represents a Hankel function of half-integral.

Our next step is to calculate the magnetic field. To do this, we will use the second of the field equations, which is given by: $b\left( {\omega ,{\bf k}} \right) = \frac{c}{v}\left[ {{k_x}{e_2}\left( {\omega ,{\bf k}} \right) - {k_y}{e_1}\left( {\omega ,{\bf k}} \right)} \right]$. Following the same procedure as outlined above, we can express the magnetic field in the form
\begin{eqnarray}
b &=& \frac{Q}{{{C_1}}}{e^{iy{\omega  \mathord{\left/
 {\vphantom {\omega  v}} \right.
 \kern-\nulldelimiterspace} v}}}\frac{{{n^2}}}{c}\int_{ - \infty }^\infty  {d{k_x}} \frac{{{k_x}}}{{\left( {k_x^2 - {\sigma ^2}} \right)}}{e^{i{k_x}x}}\nonumber\\
 &+& \frac{Q}{{{C_1}}}{e^{iy{\omega  \mathord{\left/
 {\vphantom {\omega  v}} \right.
 \kern-\nulldelimiterspace} v}}}\frac{c}{v}\frac{\omega }{{{v^2}}}\int_{ - \infty }^\infty  {d{k_x}} \frac{1}{{\left( {k_x^2 - {\sigma ^2}} \right)}}{e^{i{k_x}x}}.     \label{RadLog50}
\end{eqnarray}

After integrating over $k_{x}$, the final result for the magnetic field is therefore:
\begin{equation}
b =  - \frac{{{\pi ^{{3 \mathord{\left/
 {\vphantom {3 2}} \right.
 \kern-\nulldelimiterspace} 2}}}}}{{c\sqrt 2 }}\frac{Q}{{{C_1}}}{e^{iy{\omega  \mathord{\left/
 {\vphantom {\omega  v}} \right.
 \kern-\nulldelimiterspace} v}}}\sqrt x \left[ {{n^2}\sqrt \sigma   + \frac{{\omega {c^2}}}{{{v^3}}}\frac{1}{{\sqrt \sigma  }}} \right]H_{{1 \mathord{\left/
 {\vphantom {1 2}} \right.
 \kern-\nulldelimiterspace} 2}}^{\left( 1 \right)}\left( {\sigma x} \right). \label{RadLog55}
\end{equation}

We have finally assembled the tools to determine the radiated energy for the case under consideration. We will estimate the radiated energy by evaluating the Poynting vector ${\bf S}$. Following our earlier procedure \cite{Rad1,Rad2,Rad3}, for the case $(2+1)$-dimensional, we have
\begin{equation}
 {\bf S} = \frac{c}{{2\pi }}{\mathop{\rm Re}\nolimits} \left( {{{\bf e}_ \bot }{b^ * }} \right).  \label{RadLog65}                    
\end{equation}

To be more precise, we will compute the power radiated through the surface $S$ \cite{Das}, which reads:
 \begin{equation}
{\cal E} = \int_{ - \infty }^\infty \! {dt} \int\limits_S {d{\bf a} \cdot {\bf S}}.  \label{RadLog70}
\end{equation}

With this in view, we consider a circle with a radius of ${\rho}_{0}$. By changing to polar coordinates and taking advantage of the problem's axial symmetry, we find that the total energy radiated through the lateral surface of the circle can be expressed in the following manner:
\begin{equation}
{\cal E} = 2\pi {\rho}_{0} \int_0^\infty  {d\omega } {S_\rho }\Theta \left( {v - {c \mathord{\left/
 {\vphantom {c n}} \right.
 \kern-\nulldelimiterspace} n}} \right). \label{RadLog75a}
\end{equation}
It is noteworthy that, based on our previous discussion, we have included the step function because Cherenkov radiation is emitted only when the particle's velocity exceeds the velocity of light in the medium (${v} > \frac{c}{n}$), with $n > 1$. 

Thus, by employing equations (\ref{RadLog45}) and (\ref{RadLog55}) in the radiation zone, the expression for the power radiated per unit length reduces to: 
\begin{eqnarray}
W \equiv \frac{\cal E}{{\rho}_{0} } &=& \frac{{{\pi ^2}{Q^2}}}{{{c^2}}}\frac{v}{{{{\left( {{C_1} - {D_1}{B^2}} \right)}^2}}} \nonumber\\
&\times&\int_0^\infty  {d\omega } \left( {1 - \frac{{{c^2}}}{{{n^2}{v^2}}}} \right)\frac{1}{{\sqrt {\frac{{{n^2}{v^2}}}{{{c^2}}} - 1} }}, \label{RadLog75b}
\end{eqnarray}
where ${n^2} = \frac{{{C_1}}}{{\left( {{C_1} - {D_1}{B^2}} \right)}}$. 
This result is similar to that found in Ref. \cite{Pardy} and finds here an independent derivation.

We also observe at this stage that, since the refractive index is constant, equation (\ref{RadLog75b}) becomes
\begin{eqnarray}
W \equiv \frac{{\cal E}}{{\rho}_{0} } &=& \frac{{{\pi ^2}{Q^2}}}{{{c^2}}}\frac{v}{{{{\left( {{C_1} - {D_1}{B^2}} \right)}^2}}} \nonumber\\
&\times&\left( {1 - \frac{{{c^2}}}{{{n^2}{v^2}}}} \right)\frac{1}{{\sqrt {\frac{{{n^2}{v^2}}}{{{c^2}}} - 1} }} 
\int_0^\infty  {d\omega }, \label{RadLog75c}
\end{eqnarray}
where the integration with respect to $\omega$ is performed over the frequency range for which $n(\omega) > {c \mathord{\left/{\vphantom {c v}} \right.\kern-\nulldelimiterspace} v}$. 

However, the point we wish to emphasize is that the newly introduced vacuum can be compared to a polarizable medium, similar to the QED vacuum made up of virtual electron-positron pairs. As previously explained in our works \cite{Rad1,Rad2,Rad3}, the Cherenkov radiation produced in this framework corresponds to the energy re-emitted by excited virtual particles. In such a case, the frequency, \(\frac{2mc^2}{\hbar}\), serves as a cutoff frequency for the re-emitted photons, corresponding to the energy required for pair creation. Accordingly, we conclude that a charged particle cannot excite virtual pairs in the QED vacuum with sufficient energy for pair creation. Consequently, the re-emitted energy is constrained to \(2mc^2\), which defines the cutoff \(\Omega = \frac{2mc^2}{\hbar}\) for the frequencies considered. This frequency corresponds to the energy required for pair creation.

As a result of the last remark, equation (\ref{RadLog75c}) can be rewritten as follows:
\begin{eqnarray}
W \equiv \frac{{\cal E}}{{\rho}_{0} } &=& \frac{{{\pi ^2}{Q^2}}}{{{c^2}}}\frac{v}{{{{\left( {{C_1} - {D_1}{B^2}} \right)}^2}}} \nonumber\\
&\times&\left( {1 - \frac{{{c^2}}}{{{n^2}{v^2}}}} \right)\frac{1}{{\sqrt {\frac{{{n^2}{v^2}}}{{{c^2}}} - 1} }} 
\int_0^\Omega  {d\omega }. \label{RadLog75d}
\end{eqnarray}
In that case, we get
\begin{eqnarray}
W &=& \frac{{{2\,m\,\pi ^2}\,{Q^2}}}{{{\hbar^2}}}\frac{v}{{{{\left( {{C_1} - {D_1}{B^2}} \right)}^2}}} \nonumber\\
&\times&\left( {1 - \frac{{{c^2}}}{{{n^2}{v^2}}}} \right)\frac{1}{{\sqrt {\frac{{{n^2}{v^2}}}{{{c^2}}} - 1} }}. \label{RadLog75e}
\end{eqnarray}
We note that in the previous expression, we have omitted a constant term of the form $\sim \frac{{\pi^2 Q^2}}{{C_1^2}} \frac{{c^2}}{{v^3}} \Omega$.

\subsection{Massive nonlinear electrodynamics}

We now proceed to consider the massive case, following the procedures outlined in the previous subsection. First, we will calculate the electric field. To clarify this calculation, we will begin by presenting the field equations pertinent to the case under consideration, that is,
\begin{eqnarray}
{C_1}\nabla  \cdot {\bf e} + {M^2}\varphi  = 4\pi \,{\rho _{ext}}, \nonumber\\
\nabla  \times {\bf e} =  - \frac{1}{c}\frac{{\partial \,b}}{{\partial\, t}}, \nonumber\\
\left( {{C_1} - {D_1}{B^2}} \right)\tilde \nabla b + {M^2}\,{\bf a} = \frac{1}{c}{C_1}\frac{{\partial\, {\bf e}}}{{\partial\, t}} + 4\pi \,{{\bf J}_{ext}}, \label{RadLog80}
\end{eqnarray}
where the electric field is given by ${\bf e} =  - \nabla \varphi  - \frac{1}{c}\frac{{\partial\, {\bf a}}}{{\partial\, t}}$.

Proceeding in the same manner as in the general aspects of Section II, we obtain the following equation for the electric field:
\begin{eqnarray}
\left[ { - {\nabla ^2} + \frac{1}{{{c^2}\left( {{C_1} - {D_1}{B^2}} \right)}}\left( {{C_1}\frac{{{\partial ^2}}}{{\partial {t^2}}} + {c^2}{M^2}} \right)} \right]{\bf e}  \nonumber\\
 - \frac{{{D_1}{B^2}}}{{\left( {{C_1} - {D_1}{B^2}} \right)}}\nabla \left( {\nabla  \cdot {\bf e}} \right) =  -\, 4\pi \,\nabla {\rho _{ext}} - \frac{{4\pi }}{{{c^2}}}\frac{\partial }{{\partial t}}{{\bf J}_{ext}}. \nonumber\\
\label{RadLog85}
\end{eqnarray}
The previous equation in Fourier space can be rewritten in the following form:
\begin{eqnarray}
\left[ {\left( {{{\bf k}^2} - {n^2}\frac{{{\omega ^2}}}{{{c^2}}}} \right){\delta _{ij}} + \frac{{{D_1}{B^2}}}{{\left( {{C_1} - {D_1}{B^2}} \right)}}{k_i}{k_j}} \right]{e_j} =  \nonumber\\
 - i4\pi {k_i}{\rho _{ext}} + i4\pi \frac{\omega }{{{c^2}}}J_i^{ext}. \label{RadLog90}
\end{eqnarray}
To get the last equation, we have used ${n^2} = \frac{{{C_1} - \frac{{{M^2}{c^2}}}{{{\omega ^2}}}}}{{\left( {{C_1} - {D_1}{B^2}} \right)}}$. \\

Next, to simplify the calculation, we will assume that ${D_1}{B^2} \gg {C_1}$. In this case, equation (\ref{RadLog90}) can be expressed as: 
\begin{eqnarray}
\underbrace {\left[ {\left( {{{\bf k}^2} - {n^2}\frac{{{\omega ^2}}}{{{c^2}}}} \right){\delta _{ij}} - {k_i}{k_j}} \right]}_{{M_{ij}}}{e_j} = \nonumber\\
 - i4\pi {k_i}{\rho _{ext}} + 4\pi \frac{\omega }{{{c^2}}}J_i^{ext}.  \label{RadLog95}
\end{eqnarray}
In passing, we note that with this assumption, the refractive index becomes ${n^2} = \frac{{{M^2}{c^2}}}{{{\omega ^2}{D_1}{B^2}}}$.

We further note that the inverse of the matrix $M_{ij}$ is given by:  
\begin{equation}
M_{jl}^{ - 1} = \frac{1}{{\left( {{{\bf k}^2} - {n^2}\frac{{{\omega ^2}}}{{{c^2}}}} \right)}}{\delta _{jl}} - \frac{{{c^2}}}{{{n^2}{\omega ^2}}}\frac{1}{{\left( {{{\bf k}^2} - {n^2}\frac{{{\omega ^2}}}{{{c^2}}}} \right)}}{k_j}{k_l}. \label{RadLog100}
\end{equation}

With the aid of equations (\ref{RadLog95}) and (\ref{RadLog100}), we find that the components of the electric field can be expressed in the form
\begin{eqnarray}
{e_1} =  - i4\pi \left\{ {\frac{1}{{\left( {{{\bf k}^2} - {n^2}\frac{{{\omega ^2}}}{{{c^2}}}} \right)}} - \frac{{{c^2}}}{{{n^2}{\omega ^2}}}\frac{{k_x^2}}{{\left( {{{\bf k}^2} - {n^2}\frac{{{\omega ^2}}}{{{c^2}}}} \right)}}} \right\}{k_x}{\rho _{ext}}  \nonumber\\
 - i4\pi \left\{ { - \frac{{{c^2}}}{{{n^2}{\omega ^2}}}\frac{{{k_x}{k_y}}}{{\left( {{{\bf k}^2} - {n^2}\frac{{{\omega ^2}}}{{{c^2}}}} \right)}}} \right\}{k_y}{\rho _{ext}} \nonumber\\
 + i\frac{{4\pi }}{{{c^2}}}\omega \left\{ { - \frac{{{c^2}}}{{{n^2}{\omega ^2}}}\frac{{{k_x}{k_y}}}{{\left( {{{\bf k}^2} - {n^2}\frac{{{\omega ^2}}}{{{c^2}}}} \right)}}} \right\}{J_y}, \nonumber\\
\label{RadLog105}
\end{eqnarray}
and
\begin{eqnarray}
{e_2} =  - i4\pi \left\{ { - \frac{{{c^2}}}{{{n^2}{\omega ^2}}}\frac{{{k_x}{k_y}}}{{\left( {{{\bf k}^2} - {n^2}\frac{{{\omega ^2}}}{{{c^2}}}} \right)}}} \right\}{k_x}{\rho _{ext}} \nonumber\\
 - i4\pi \left\{ {\frac{1}{{\left( {{{\bf k}^2} - {n^2}\frac{{{\omega ^2}}}{{{c^2}}}} \right)}} - \frac{{{c^2}}}{{{n^2}{\omega ^2}}}\frac{{k_y^2}}{{\left( {{{\bf k}^2} - {n^2}\frac{{{\omega ^2}}}{{{c^2}}}} \right)}}} \right\}{k_y}{\rho _{ext}} \nonumber\\
 + \frac{{i4\pi }}{{{c^2}}}\omega \left\{ {\frac{1}{{\left( {{{\bf k}^2} - {n^2}\frac{{{\omega ^2}}}{{{c^2}}}} \right)}} - \frac{{{c^2}}}{{{n^2}{\omega ^2}}}\frac{{k_y^2}}{{\left( {{{\bf k}^2} - {n^2}\frac{{{\omega ^2}}}{{{c^2}}}} \right)}}} \right\}{J_y}. \nonumber\\
\label{RadLog110}
\end{eqnarray}

Again, making use of ${\rho _{ext}} = 2\pi Q\delta \left( {\omega  - {k_y}v} \right)$ and ${{\bf J}_{ext}} = 2\pi Qv\delta \left( {\omega  - {k_y}v} \right)\pmb {\hat e_y}$, we readily find that
the electric field components can be written as follows
\begin{eqnarray}
{e_1}\left( {\omega ,x} \right) = i\frac{{2Q}}{v}{e^{i{{y\omega } \mathord{\left/
 {\vphantom {{y\omega } v}} \right.
 \kern-\nulldelimiterspace} v}}}\int_{ - \infty }^\infty  {d{k_x}} {e^{i{k_x}x}}\left\{ {\frac{{{c^2}}}{{{n^2}{v^2}}}\frac{{{k_x}}}{{\left( {k_x^2 - {\sigma ^2}} \right)}}} \right\} \nonumber\\
 + i\frac{{2Q}}{v}{e^{i{{y\omega } \mathord{\left/
 {\vphantom {{y\omega } v}} \right.
 \kern-\nulldelimiterspace} v}}}\int_{ - \infty }^\infty  {d{k_x}} {e^{i{k_x}x}}\left\{ { - \left( {1 + \frac{1}{{{n^2}}}} \right)\frac{{{k_x}}}{{\left( {k_x^2 - {\sigma ^2}} \right)}}} \right\} \nonumber\\
 + i\frac{{2Q}}{v}{e^{i{{y\omega } \mathord{\left/
 {\vphantom {{y\omega } v}} \right.
 \kern-\nulldelimiterspace} v}}}\int_{ - \infty }^\infty  {d{k_x}} {e^{i{k_x}x}}\left\{ {\frac{{{c^2}}}{{{n^2}{\omega ^2}}}\frac{{k_x^3}}{{\left( {k_x^2 - {\sigma ^2}} \right)}}} \right\}, \nonumber\\
\label{RadLog115}
\end{eqnarray}
and
\begin{eqnarray}
{e_2}\left( {\omega ,x} \right) = i\frac{{2Q}}{v}{e^{i{{y\omega } \mathord{\left/
 {\vphantom {{y\omega } v}} \right.
 \kern-\nulldelimiterspace} v}}}\int_{ - \infty }^\infty  {d{k_x}} {e^{i{k_x}x}}\left\{ {\frac{{{c^2}}}{{{n^2}{v^2}}}\frac{{k_x^2}}{{\left( {k_x^2 - {\sigma ^2}} \right)}}} \right\} \nonumber\\
 + i\frac{{2Q}}{v}{e^{i{{y\omega } \mathord{\left/
 {\vphantom {{y\omega } v}} \right.
 \kern-\nulldelimiterspace} v}}}\int_{ - \infty }^\infty  {d{k_x}} {e^{i{k_x}x}} \nonumber\\
 \times \left\{ {\frac{{\omega v}}{{{c^2}}}\left( {1 - \frac{{{c^2}}}{{{v^2}}}} \right)\left( {1 - \frac{{{c^2}}}{{{n^2}{v^2}}}} \right)\frac{1}{{\left( {k_x^2 - {\sigma ^2}} \right)}}} \right\}. \nonumber\\
\label{RadLog120}
\end{eqnarray}

In passing, the integrals over $k_{x}$ are identical to those in the previous subsection. In this manner, we obtain the following expression for the electric field:
\begin{eqnarray}
{\bf e}\left( {\omega ,{\bf x}} \right) = i{\pi ^{{3 \mathord{\left/
 {\vphantom {3 2}} \right.
 \kern-\nulldelimiterspace} 2}}}\frac{{2Q}}{v}{e^{iy{\omega  \mathord{\left/
 {\vphantom {\omega  v}} \right.
 \kern-\nulldelimiterspace} v}}} \nonumber\\
\times \sqrt {\frac{{\sigma x}}{2}} \left[ {\left( {1 - \frac{{{c^2}}}{{{n^2}{v^2}}}} \right) + \frac{1}{{{n^2}}} - \frac{{{c^2}}}{{{n^2}{\omega ^2}}}{\sigma ^2}} \right]H_{{1 \mathord{\left/
 {\vphantom {1 2}} \right.
 \kern-\nulldelimiterspace} 2}}^{\left( 1 \right)}\left( {\sigma x} \right)\pmb {{\hat e}_x} \nonumber\\
 + i{\pi ^{{3 \mathord{\left/
 {\vphantom {3 2}} \right.
 \kern-\nulldelimiterspace} 2}}}\frac{{2Q}}{v}{e^{iy{\omega  \mathord{\left/
 {\vphantom {\omega  v}} \right.
 \kern-\nulldelimiterspace} v}}} \nonumber\\
\times \sqrt {\frac{{\sigma x}}{2}} \left[ { - \frac{{\omega v}}{{{c^2}}}\left( {1 - \frac{{{c^2}}}{{{v^2}}}} \right)\left( {1 - \frac{{{c^2}}}{{{n^2}{v^2}}}} \right) - \frac{{{c^2}}}{{{n^2}\omega v}}\sigma } \right] \nonumber\\
\times H_{{1 \mathord{\left/
 {\vphantom {1 2}} \right.
 \kern-\nulldelimiterspace} 2}}^{\left( 1 \right)}\left( {\sigma x} \right) \pmb{{\hat e}_y}. 
\nonumber\\
\label{RadLog125}
\end{eqnarray}

It may be noticed here that by repeating the same steps that led to equation (\ref{RadLog55}), the magnetic field has the form
\begin{equation}
b =  - i\frac{{2Q}}{c}{\pi ^{{3 \mathord{\left/
 {\vphantom {3 2}} \right.
 \kern-\nulldelimiterspace} 2}}}{e^{i{{y\omega } \mathord{\left/
 {\vphantom {{y\omega } v}} \right.
 \kern-\nulldelimiterspace} v}}}\sqrt {\frac{{\sigma x}}{2}} H_{{1 \mathord{\left/
 {\vphantom {1 2}} \right.
 \kern-\nulldelimiterspace} 2}}^{\left( 1 \right)}\left( {\sigma x} \right). \label{RadLog130}
  \end{equation}
  
In conclusion, by utilizing equations (\ref{RadLog65}), (\ref{RadLog70}), (\ref{RadLog75a}), (\ref{RadLog125}) and (\ref{RadLog130}) within the radiation zone, we find that, following additional manipulations, the power radiated per unit length can be expressed in the form 
\begin{eqnarray}
W \equiv \frac{{\cal E}}{\rho } &=& \frac{{2{\pi ^2}{Q^2}}}{c}\left( {1 - \frac{{{c^2}}}{{{v^2}}}} \right) \nonumber\\
&\times& \int_0^\infty  {d\omega } \left( {1 - \frac{{{c^2}}}{{{n^2}{v^2}}}} \right)\frac{1}{{\sqrt {\frac{{{n^2}{v^2}}}{{{c^2}}} - 1} }}.  \label{RadLog135}
\end{eqnarray}

Following our earlier discussion, we observe that the power radiated per unit length is expressed as:
\begin{eqnarray}
W \equiv \frac{{\cal E}}{\rho } &=& \frac{{2{\pi ^2}{Q^2}}}{c}\left( {1 - \frac{{{c^2}}}{{{v^2}}}} \right) \nonumber\\
&\times& \int_0^\Omega  {d\omega } \left( {1 - \frac{{{c^2}}}{{{n^2}{v^2}}}} \right)\frac{1}{{\sqrt {\frac{{{n^2}{v^2}}}{{{c^2}}} - 1} }}.  \label{RadLog140}
\end{eqnarray}
We also note that, similar to the massless case, we have omitted a term $\sim  - \frac{{2\sqrt 2 c}}{{\pi v}}\frac{1}{{\sqrt \rho  }}\int_0^\infty  {d\omega } \frac{1}{n}\sqrt {1 - \frac{{{c^2}}}{{{n^2}{v^2}}}}$, which does not contribute in the far zone $\rho  \to \infty$. \\

\section{Final remarks}

This work presents another observational signature of electromagnetic field non-linearities in $(1+2)$ dimensions based on the standard Poynting vector calculation. 
Specifically, we have considered the electromagnetic radiation emitted by a moving charged particle interacting with a medium for massless and massive nonlinear electrodynamics.

In conclusion, we would like to emphasize that, although it was not the scope of this contribution, we have examined the complete Maxwell-Chern-Simons action, incorporating non-linear terms and a de Broglie-Proca mass term in Fourier space. By employing specific properties of the curl and divergence operators in two Euclidean dimensions, it was possible to define a dielectric tensor in the vacuum with a wave-vector dependence profile. This means that the vacuum of the full action under investigation behaves like a non-local material (in the sense of a spatially-dispersed profile). We understand that this phenomenon is a consequence of the non-vanishing de Broglie-Proca mass. However, this is a subtle point that deserves further explanation. If we return to Eq. (\ref{Fieldeq70}), it is evident that the terms bilinear in the wave vector, which responds to the non-linear material behavior, appear with the topological mass, $m$, and not with the mass parameter, $M^2$. In the intermediate steps between Eqs. \eqref{Fieldeq30},\eqref{Fieldeq35},\eqref{Fieldeq40} and Eq.(\ref{Fieldeq70}), we have assumed that $M^2$ is non-vanishing, which allows us to divide by this parameter. However, when considering the Amp\`ere-Maxwell (\ref{Fieldeq40}), we encounter a multiplication by $M^2$. As an immediate consequence, this operation yields the presence of the topological mass, m, coupled to the wave vector components, while $M^2$ disappears from the ${\bf k_{i}}$-dependent terms in (\ref{Fieldeq70}). We have also checked that, by considering only the Chern-Simons mass and switching off the $M^2$-parameter, the spatial dispersion profile does not show up; all we get is a dielectric tensor that depends only on the frequency. Therefore, we may state that the de Broglie-Proca mass term is responsible for the non-local material/metamaterial behavior of the complete three-dimensional electrodynamics we have reported here. It remains to be inspected, as demonstrated by Mikki, in $(1+3)$- dimensions \cite{Mikki3}, if the purely Maxwellian theory in a planar non-local metamaterial is equivalent to the $3$D Maxwell-de Broglie-Proca theory in vacuum. 
We shall report on that elsewhere.

\section*{Acknowledgments}

%One of us (P. G.) was partially supported by ANID PIA / APOYO AFB220004.
One of us (P. G.) acknowledges the financial support received by ANID PIA/APOYO AFB230003.

\end{document}